\documentclass[%
reprint,
showkeys,
amsmath,amssymb,
aps,
pre,
]{revtex4-1}

\usepackage{amsmath}
\usepackage{multirow}
\usepackage{subfigure}
\usepackage{hyperref}

\usepackage{lineno}
\modulolinenumbers[5]

\usepackage{amssymb}	
\newcommand{\nnn}[2]{\eta_{{#1}{#2}}}
\newcommand{\nk}[1]{\nnn{#1}{\checkmark}}	
\newcommand{\nn}[1]{\nnn{#1}{\times}}  		
\newcommand{\numE}{\nnn{}{\times}}  		
\newcommand{\accuracyE}{\delta_{\times}}		

\newcommand{\reffig}[1]{Fig.~\ref{#1}}
\newcommand{\reftbl}[1]{Table~\ref{#1}}

\newcommand{\refsec}[1]{Sec.~\ref{#1}}

\usepackage{graphicx}

\begin{document}

\preprint{APS/123-QED}

\title{Getting recommendation is not always better}

\author{Zeynep~B.~Cinar}
\email{zeynepbusracinar@gmail.com}
\author{Haluk~O.~Bingol}%
\email{bingol@boun.edu.tr}
\affiliation{%
	Department of Computer Engineering, 
	Bogazici University, 
	Istanbul,
	34342 Turkey \\
}%

\begin{abstract}
We present an extended version of the Iterated Prisoner's Dilemma game
in which agents with limited memory receive recommendations 
about the unknown opponents to decide whether to play with.
Since agents can receive more than one recommendation about the same opponent,
they have to evaluate the recommendations according to their disposition
such as optimist, pessimist, or realist.
They keep their firsthand experience in their memory.
Since agents have limited memory,
they have to use different forgetting strategies.
Our results show that getting recommendations does not always perform better.
With the support of recommendation, cooperators can beat defectors.
We observe that realist performs the best and optimist the worse.
\end{abstract}


\maketitle

\section{Introduction}

We live in a complex and crowded world.
It is so crowded that it is impossible to ``know'' everybody~\cite{bingol2008fame}.
We know only a very small fraction of the population.
We operate within this network of known people.
As we interact with them, 
we classify them based on our firsthand experience.
Next time we need to interact again,
we use this information.

Since bold interaction with an unknown person may not be a good idea,
proceeding with firsthand experience is not the only way 
that we use
to make a judgement about somebody.
If we do not know a person,
we ask for recommendations from people that we already know. 
We do that all the time in real life.

It is common sense that having recommendation is better.
In this work, we investigate whether having a recommendation is always a better alternative
in the context of iterated prisoner's dilemma game. 
Our findings indicate that it is not always the case.

\section{Background}

\subsection{Optimism and pessimism}
\label{sec:optimismPessimism}

There are different orientations in humans to suggest that what kind of things they could do in a given situation~\cite{%
	susan1991dynamics}.
These orientations have an impact on trust~\cite{%
	sherchan2013survey}.
Marsh presents dispositions in terms of how an agent estimates trust~\cite{%
	marsh1994optimism}.
Optimism, pessimism and realism are the notions of dispositions of trust.
Each disposition results in different trust estimations.
An ``optimist'' expects the best in people and is always hopeful about the result of the situations.
A ``pessimist'', unlike the optimist, sees the worst in people, always looks at the situations through doubting eyes.
While optimism and pessimism are extreme cases,
there are other cases such as ``realism''.

\subsection{Prisoner’s dilemma}

In prisoner’s dilemma, agents ``cooperate'' or ``defect'', without knowing their opponent’s choice~\cite{%
	axelrod1981evolution}. 
As summarized in \reftbl{tab:payoffMatrix},
if an agent cooperates while the other defects, cooperator gets the sucker payoff $S$ and defector gets the temptation payoff $T$. 
On the other hand, 
if both players choose to cooperate, 
then they both get the reward payoff $R$. 
Lastly, 
in the case of mutual defection, both players get the punishment payoff $P$. 
In Prisoner’s Dilemma game, 
the payoffs should satisfy both $S < P < R < T$ and $S + T < 2 R$.

\begin{table}[ph]
\begin{center}
	\caption{%
		\label{tab:payoffMatrix}%
		Payoff matrix. $(S, P, R, T) = (0, 1, 3, 5)$.
	}
	\begin{tabular}{*{4}{c|}}
		\multicolumn{2}{c}{} &\multicolumn{2}{c}{Player $X$}\\
		\cline{3-4}
		\multicolumn{1}{c}{} &  &Cooperate  &Defect\\
		\cline{2-4}
		\multirow{2}*{Player $Y$}  &Cooperate &$(R,R)$ &$(S,T)$\\
		\cline{2-4}
		&Defect &$(T,S)$ &$(P,P)$\\
		\cline{2-4}
	\end{tabular}
	\label{table:payoffMatrix}
\end{center}
\end{table}

For a single Prisoner's Dilemma game, it is more advantageous to defect.
But, when the game repeats, things change.
Iterated Prisoner's Dilemma (IPD) differs from the original concept of a prisoner's dilemma
because players can remember the past interactions of their opponent and can change their strategy accordingly~\cite{axelrod1981evolution,Leas2016prisoners}.
Players can learn about the behaviors of their opponent and have the opportunity to penalize the agents for previous defective decisions.

Iterated prisoner's dilemma is used to understand cooperation.
Some evolutionary approaches, 
which promotes cooperation,
are investigated in the literature~\cite{nowak2006five,chu2017win,yamamoto2019effect}.
Some works based on the reputation of the agents,
where agents cooperate with the ones that have good reputation~\cite{axelrod1981evolution,nowak1998dynamics,nowak1998evolution}.
In this paper, we propose a non-evolutionary model that also promotes cooperators.
We will see that cooperators can get the average payoff that is larger than that of defectors.

\subsection{Iterated prisoner's dilemma with limited attention}

In ``Iterated prisoner's dilemma with limited attention'' (IPDwLA) model, 
agents play Iterated Prisoner's Dilemma game 
in which players can accept or refuse to play with their partner~\cite{Cetin2014,Cetin2016}.
There are $N$ agents in the population.
Agents are not pure cooperators or pure defectors.
Agent $i$ cooperates with probability $\rho_{i}$, 
which is called \emph{cooperation probability}.
There are two types of agents,
one group, called \emph{cooperators},  has a $\rho$ value larger than $0.5$.
The other group, called \emph{defectors}, has a $\rho$ value, which is less than $0.5$.

\subsubsection{Decision to play}

IPDwLA adopts choice-and-refusal rule~\cite{%
	stanley1993iterated}: 
If an agent ``knows'' that the opponent is a defector, then it refuses to play. 
Otherwise, it plays.
That is,
two agents are randomly selected and offered to play Iterated Prisoner's Dilemma game.
Both agents evaluate their opponent and decide whether to play or not.
(i)~If an agent does not know the opponent, then it has to play.
(ii)~If it ``knows'' the opponent as ``cooperator'', then it plays.
(iii)~If $i$ ``knows'' $j$ as ``defector'', then $i$ refuses to play.
Note that a game takes place 
only if both players decide to play.

\subsubsection{Perception}

In order agents to ``know'' each other, 
agents have some memory so that they can keep track of previous games with the same agent.
Suppose agent $i$ plays with agent $j$.
Agent $i$ keeps two numbers in its memory.
The number $c_{ji}$ is the number of times that $j$ cooperates and
$d_{ji}$ is the number of times that $j$ defects
when $j$ plays with $i$.
Then the \emph{perceived cooperation ratio} is defined as
\begin{equation}
	t_{i j} = c_{ji} / ( c_{ji} + d_{ji} ).
	\label{eq:perceivedCooperationRatioCetin}
\end{equation}
If the perceived cooperation ratio is larger than $0.5$,
$i$ considers $j$ as \emph{cooperator}, otherwise as \emph{defector}.

\subsubsection{Memory}

We assume that each agent has an identical memory capacity of size $M \le N$,
called \emph{memory size}.
That is, each agent can keep track of at most $M$ opponents.
\emph{Memory ratio}, defined as $\mu = M / N \in [0,1]$, 
is the percentage of agents that can be kept in one's memory.

\subsubsection{Forgetting strategies}

Agent stores each opponent in a different slot in its memory.
Eventually, the agent will run out of memory for $M < N$.
After that point, to create memory space for a new opponent, 
the agent has to ``forget'' a known opponent.
There are several forgetting mechanisms investigated in the model.
(i)~Players prefer to forget cooperators first, denoted by FC. 
(ii)~Players prefer to forget defectors first, denoted by FD.
(iii)~Players prefer to forget randomly, denoted by FR~\cite{%
	Cetin2014}.

\section{Proposed model (IPDwRec)}

In IPDwLA model, if an agent does not know the opponent,
it has to play~\cite{Cetin2014,Cetin2016}.
However, in real life,
we use our social network to obtain information about a person
that we do not know.
We extend the model 
so that agents can get recommendations about the opponents that they do not know.

\begin{figure}[htbp]
\centering
	\includegraphics[width=.4\columnwidth]%
		{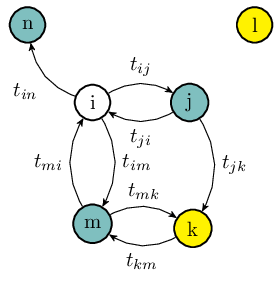}
	\caption{
		Perception graph is a weighted directed graph,
		where vertices represent agents.
		There is a directed edge from agent $i$ to agent $j$,
		if $i$ previously played with $j$.
		The weight $t_{ij}$ represents the perceived cooperation ratio of $j$ 
		with respect to $i$.
		Agents $j$, $m$ and $n$ are ``known'' by $i$,
		while $\ell$ and $k$ are not.
	}
	\label{fig:perceptionGraph}
\end{figure}

\subsection{Perception graph}

Perceptions between agents can be represented as a weighted directed graph 
as given in \reffig{fig:perceptionGraph}.
In a \emph{perception graph}, 
agents are represented by vertices.
When $i$ plays with $j$ for the first time, 
two arcs will be created, namely,
one from $i$ to $j$,
and one from $j$ to $i$. 
The perceived cooperation ratio of $j$ with respect to $i$, denoted by $t_{ij}$, is assigned to the arc from $i$ to $j$ as a weight. 

As the game proceeds,
symmetric connectivity between agents can break
due to forgetting mechanism,
as in the case of $j$ and $k$.
Suppose agents $j$ and $k$ played before.
Hence arcs $(j, k)$ and $(k, j)$ were created.
Later, when agent $k$ chooses to forget $j$ due to lack of memory space,
arc $(k, j)$ and its corresponding weight $t_{kj}$ are removed from the graph and its memory.
Note that since $j$ still keeps $k$ in its memory, arc $(j, k)$ and $t_{jk}$ are intact.

\subsection{Decision to play or not}

In \reffig{fig:perceptionGraph}, 
suppose $i$ is one of the selected agents to play.
As far as the opponent is concerned, 
there are three cases.\\
(i)~\emph{Known opponent.}
For example,
$j$ is known by $i$.
Then all $i$ has to do is to check the perceived cooperation ratio.
If $t_{ij} > 0.5$, it plays.

However, 
if there is no data about the opponent in its memory, 
as in the cases of $k$ or $\ell$, 
agent $i$ plays with it in IPDwLA model~\cite{%
	Cetin2016}.
In our model, 
IPDwRec,
agent $i$ asks for ``recommendations'' from its neighbors.	
There are two possible cases:\\
(ii)~\emph{Unknown opponent without any recommendation.}
If nobody in its neighborhood knows the opponent, 
$i$ plays with the opponent as in the case of $\ell$.\\
(iii)~\emph{Unknown opponent with recommendations.}
Any neighbor, such as $j$, that knows $k$, 
provides its own perception $t_{j k}$ about $k$.
If there was only one such agent $j$, decision of $i$ would be relatively easy: 
it will play if $t_{j k} > 0.5$.
But usually, there are many such agents, 
e.g. $j$ and $m$.
Then $i$ has to evaluate conflicting recommendations received from them.

\subsection{Evaluation of recommendations}

We define \emph{1-neighborhood} of $i$, 
denoted by $\Gamma(i)$, 
as all agents to which there is an arc from $i$. 
Note that $\Gamma(i)$ is composed of agents in $i$'s memory. 
Hence if an agent $j$ is removed from $i$'s memory, 
it is also removed from $\Gamma(i)$. 

The set of recommender agents of $i$ about $k$ is denoted by
$
	R_{i k} 
	= 
	\{
		j \in \Gamma(i)
		\mid 
		k \in \Gamma(j)
	\}.
$ 
For example, in \reffig{fig:perceptionGraph}, $R_{i k} = \{j , m\}$.
Note that, $n$ is not in $R_{i k}$ since $n$ does not know $k$.    
Every agent in $R_{i k}$ gives a recommendation to $i$.

Once the agent receives the recommendations, the evaluation process begins.
Evaluation of the recommendations by agents varies according to their dispositions.
We consider three types of dispositions:   \\
(i)~\emph{Optimists.} 
An optimist agent $i$ takes the maximum of the recommendations that it receives~\cite{marsh1994optimism},
that is,
\[
	t_{ik} =  \max\{t_{j k}  \mid  j \in R_{i k}\}.
\]
(ii)~\emph{Pessimists.}
A pessimist agent $i$ takes the minimum of the recommendations that it gets~\cite{marsh1994optimism},
namely,
\[
	t_{ik} = \min\{t_{j k} \mid j \in R_{i k}\}.
\]
Optimist and pessimist agents are the opposite of each other.
We consider a third type in between. \\
(iii)~\emph{Realists.}
A realist agent $i$ takes the average of the recommendations~\cite{marsh1994optimism},
that is,
\[
	t_{ik} 
	= 
	\frac{1}{| R_{i k} |} 
	\sum_{j \in R_{i k}} {t_{j k}},
\]
where $|A|$ denotes the cardinality of set $A$.
In the literature, 
there are realist agents 
that use the mode or the median of the recommendations, too~\cite{%
	Marsh1994formalisingtrust}.    \\
(iv)~\emph{Self-assured (SA).}
To compare our agents with the previous work~\cite{%
	Cetin2014,%
	Cetin2016},
we also consider agents that do not ask for recommendations.
They use their memory only.

\subsection{Perceived cooperation ratio}

We modify the definition of perceived cooperation ratio of IPDwLA.
According to Eq.~(\ref{eq:perceivedCooperationRatioCetin}),
if agent $j$ plays with $i$ once and cooperates, 
$t_{i j}$ will be $1$.
Similarly, 
if $j$ plays with $i$, 
say 10 times, 
and cooperates in all of them, 
$t_{i j}$ is still equal to $1$.
However, in real life, 
the trustworthiness of a person who was honest with us only once and 
that of a person who was honest with us many times are not the same~\cite{%
	kiefhaber2013ranking}.
With this idea, 
we propose a different approach to calculations of perceived cooperation ratio.
\emph{Perceived cooperation ratio} of $j$ with respect to $i$ is defined as

 \begin{equation}
	t_{i j} = 
		\frac
			{c_{ji} + 1}
			{(c_{ji} + 1) + (d_{ji} + 1)}.
	\label{eq:perceivedCooperationRatio}
\end{equation}
We would like to note that this is one form of Laplace's rule of success~\cite{jaynes2003probability,zabell1989rule}.

With our new formula, 
if $j$ plays with $i$ once and cooperates, 
$t_{i j}$ becomes $0.66$. 
On the other hand, 
if $j$ plays with $i$ 10 times and cooperates in all of them, 
$t_{i j}$ becomes $0.91$. 
Hence Eq.~(\ref{eq:perceivedCooperationRatio})
provides more realistic perception evaluation.
Note that both 
Eq.~(\ref{eq:perceivedCooperationRatioCetin}) and 
Eq.~(\ref{eq:perceivedCooperationRatio})
produce the same value of $t_{ij} = 0.5$  when $c_{ji} = d_{ji}$.
Therefore, 
since the maximum or the minimum are still either larger or smaller than 0.5,
optimist and pessimist agents are not affected
from this new definition.
On the other hand, 
realist agents are affected 
since average will be different.

\subsection{Metrics}

\subsubsection{Payoff ratio}

We want to compare the performances of the cooperators and the defectors at the end of the game.
Let $\mathcal{A}$ be a set of agents.
Then we define \emph{average payoff} of agents in $\mathcal{A}$ as
\begin{equation}
	\overline{P_{\mathcal{A}}} 
	= 
	\frac{1}{| \mathcal{A} |}
		\sum_{i \in \mathcal{A}} \text{payoff}(i).
\end{equation}
where
$\text{payoff}(i)$ denotes the accumulated payoffs by agent $i$ at the end of the game.
%

We have two sets of agents.
The set $\mathcal{C}$ of agents with cooperation probability $\rho > 0.5$ are considered cooperators.
The rest of the agents are called defectors and denoted by $\mathcal{D}$.
Now, we can define \emph{payoff ratio} as 
\begin{equation}
	\phi_{C} = 
		\overline{P_{\mathcal{C}}} 
		\ / \ 
		\overline{P_{\mathcal{C} \cup \mathcal{D}}}.
\end{equation}
Note that we are interested in cases of $\phi_{C} > 1$, 
where cooperators have higher average payoff than the average,
that is, they perform better than defectors.
See \reffig{fig:FC_PayoffRatio}.

\subsubsection{Accuracy}

Agents can misjudge their opponents in two ways.
A cooperator is considered as defector
or a defector is considered as cooperator.
They are denoted by $cd$ and $dc$, respectively.
The correct judgments are represented by $cc$ and $dd$.
The process of judgment is done with and without memory.
(i)~They use their ``perception''
if they know the opponent in their memory.
(ii)~If they do not know the opponent,
then they ask for recommendations and ``evaluate'' what they received.
Symbols $\checkmark$ and $\times$ indicate the first and the second cases, respectively.
See \reffig{fig:FC_eval_percept} for all possible combinations of perceptions and evaluations.
For example, 
the number of failures to perceive a cooperator as defector is denoted as $\nk{cd}$.
Similarly, $\nn{dc}$ is the number of evaluations of a defector as cooperator.
\emph{Evaluation accuracy} is defined as 
\begin{equation}
	\accuracyE = 1 - \frac
						{\nn{cd} + \nn{dc}}
						{\eta_{\times}},
\end{equation}
where
$\eta_{\times}$ is the total number of recommendations evaluated.
Clearly, we have
$\eta_{\times} = \nn{cc} + \nn{dd} + \nn{cd} + \nn{dc}$. 
Note that it is possible that nobody in the 1-neigborhood  knows the opponent.
Then no recommendation is received to evaluate.

\section{Results}

Given the model,
it is possible to come up with many possible scenarios.
In this paper, we investigate some of them.
(i)~We consider homogeneous agents, 
i.e.,
one type of cooperators with cooperation probability $\rho = 0.9$ playing against one type of defectors with $\rho = 0.1$.
(ii)~In our experiments, 
both cooperators and defectors use the same forget strategy such as forget cooperators first (FC).
(iii)~Defectors are always \emph{self-assured},
i.e. they do not ask for a recommendation.

In contrast to defectors,
cooperators do ask for a recommendation.
Therefore,
for a given parameter set,
an experiment consists of four different simulations
where cooperators are either
(i)~optimistic or
(ii)~pessimistic or
(iii)~realist or
(iv)~self-assured
playing against self-assured defectors.
We will be using plots such as \reffig{fig:FC_PayoffRatio} 
to compare performances of different dispositions.
We run experiments for various memory ratios and 
report our findings as a function of the memory ratio of $\mu$.

In each experiment,
there are $N = 100$ agents,
where 50 defectors play against 50 cooperators,
i.e., 
$50~\%$ cooperators.
As in the case of IPDwLA model~\cite{%
	Cetin2016},
we terminate the experiments after $\tau {N \choose 2}$ pairs invited to play, 
where $\tau = 30$.
That is,
any pair of $N$ agents have a chance to play 30 times on average.
Note that having invited to play does not necessarily mean that the two will play.
We use traditional payoff matrix of $(S, P,  R, T) = (0, 1, 3, 5)$~\cite{%
	axelrod1981evolution}.
We report an average of 50 realizations.

\begin{figure}[htpb]
\centering
		\includegraphics[width= 0.5\columnwidth]%
			{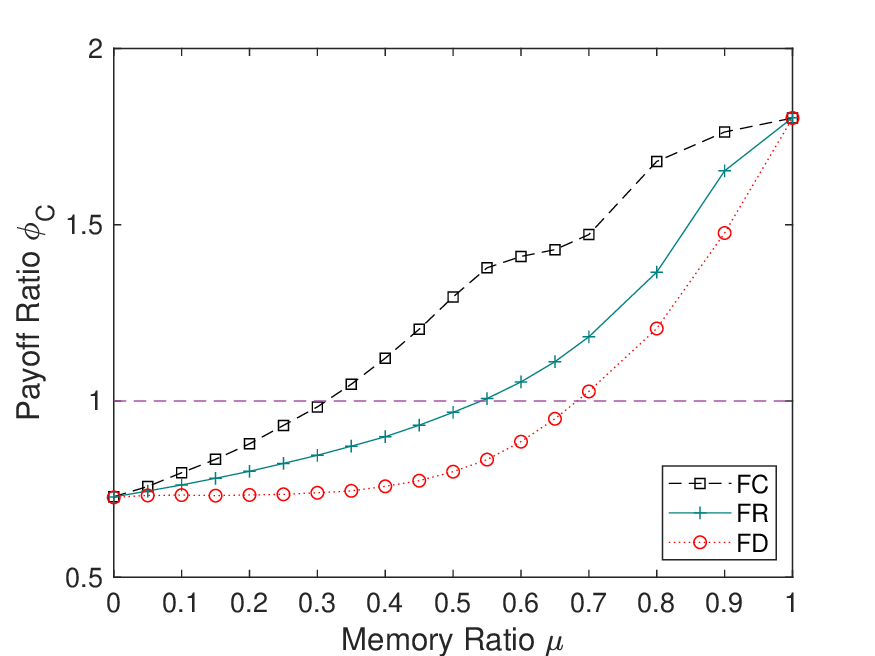}
	\caption{
		Comparison of forget mechanisms for self-assured agents.
		FC - Forget Cooperators,
		FR - Forget Randomly,
		FD - Forget Defectors.
	}
	\label{fig:SA_All}
\end{figure}

\subsection{Self-assured agents}

First of all,
we consider agents that do not get recommendations as in the case of IPDwLA model~\cite{%
	Cetin2014%
}.
\reffig{fig:SA_All} agrees with the finding of Ref~\cite{%
	Cetin2014%
}
that forgetting cooperators first (FC) is a better strategy compared to random forgetting (FR) or forgetting defectors first (FD).

An agent, 
who does not use its ``network'', 
would have to rely on its own memory only.
Therefore,
it requires ``considerable memory''.
Even FC strategy can cross the line of $\phi_{C} = 1$
only after having memory size that can hold $30~\%$ of the population ($\mu \sim 0.3$).
Strategies 
FR 
and 
FD
call for memory ratios more than $0.5$ for $\phi_{C} > 1$.

\subsection{Forget cooperators first (FC)}

We investigate forgetting cooperators first strategy 
since it is better compared to FR and FD strategies according to \reffig{fig:SA_All} and Ref~\cite{%
	Cetin2014}.
Cooperators of a disposition,
such as optimist,
play against self-assured defectors.
Both cooperators and defectors use \emph{forget cooperators first} (FC) strategy.
That is,
if there is no space left in the agent's memory,
a cooperator in the memory is randomly selected and forgotten.
If there is no cooperator left in the memory,
then a randomly selected defector is forgotten.

\begin{figure}[htpb]
	\subfigure[Payoff ratios.]{%
		\includegraphics[width=0.45\columnwidth]%
			{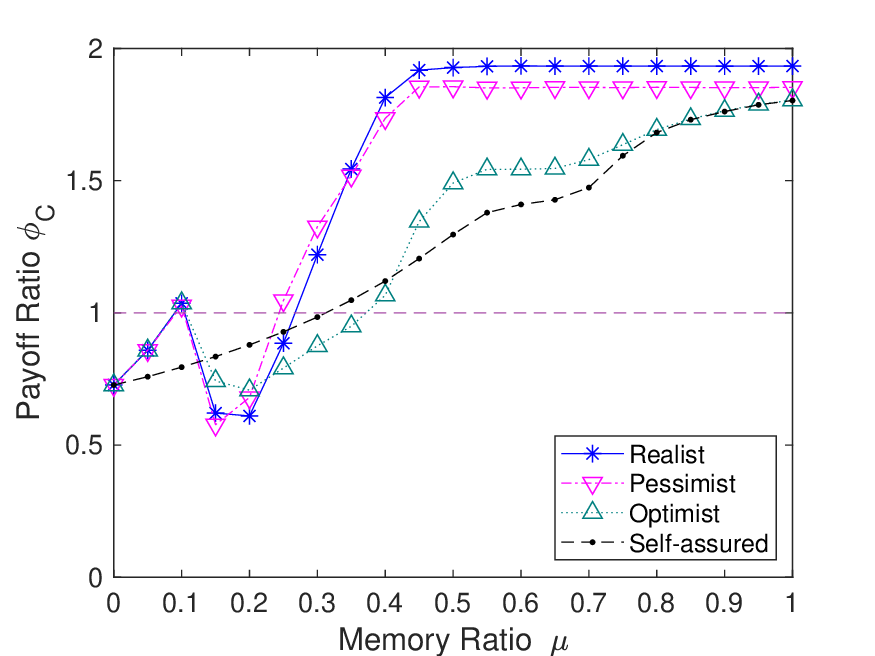}
		\label{fig:FC_PayoffRatio}%
	}%
	\subfigure[Accuracy of evaluations.]{%
		\includegraphics[width=0.45\columnwidth]%
			{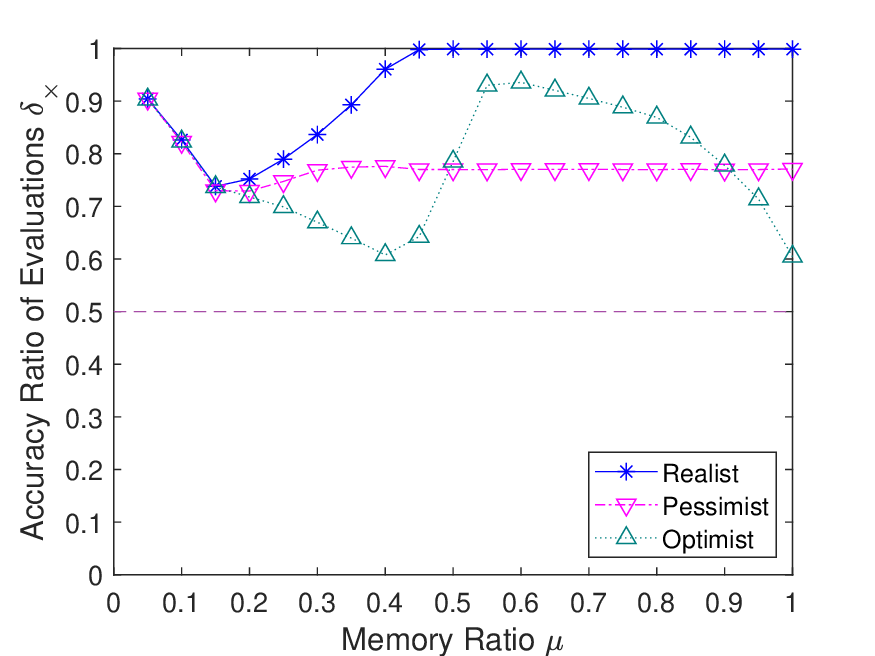}
		\label{fig:FC_AccuracyRatioEvaluation}%
	}%
	\\
	\subfigure[Number of evaluations.]{%
		\includegraphics[width=0.45\columnwidth]%
			{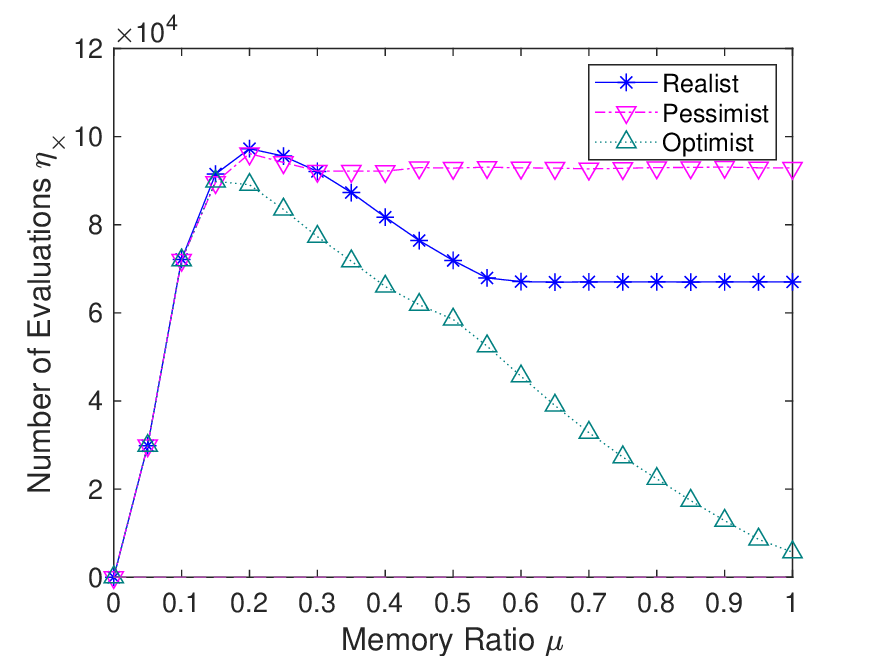}
		\label{fig:FC_NORecommendations}%
	}%
	\subfigure[Total memory usage.]{%
		\includegraphics[width=0.45\columnwidth]%
			{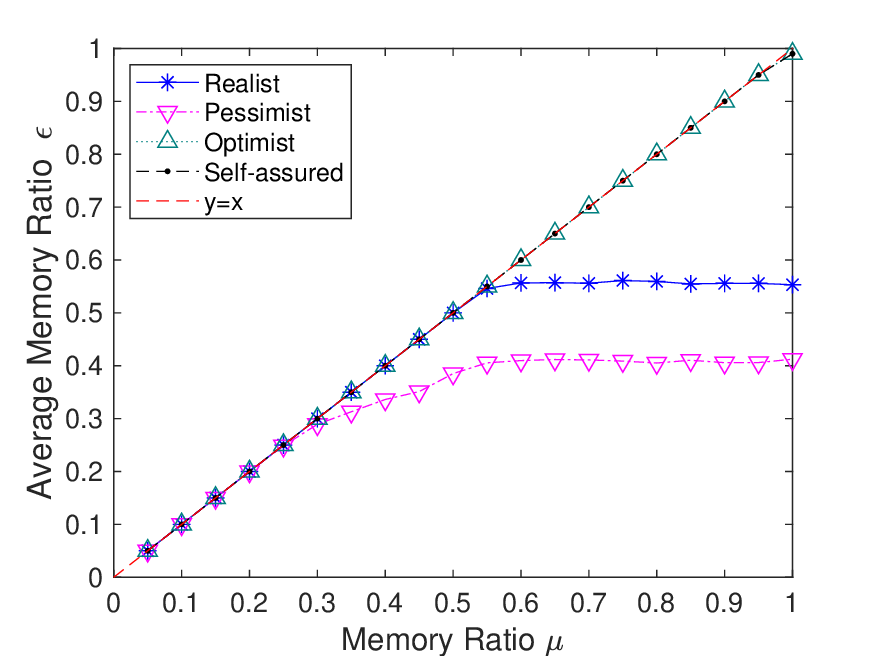}
		\label{fig:FC_memoryUsage_Total}%
	}%
	\\
	\subfigure[Cooperators in memory.]{%
		\includegraphics[width=0.45\columnwidth]%
			{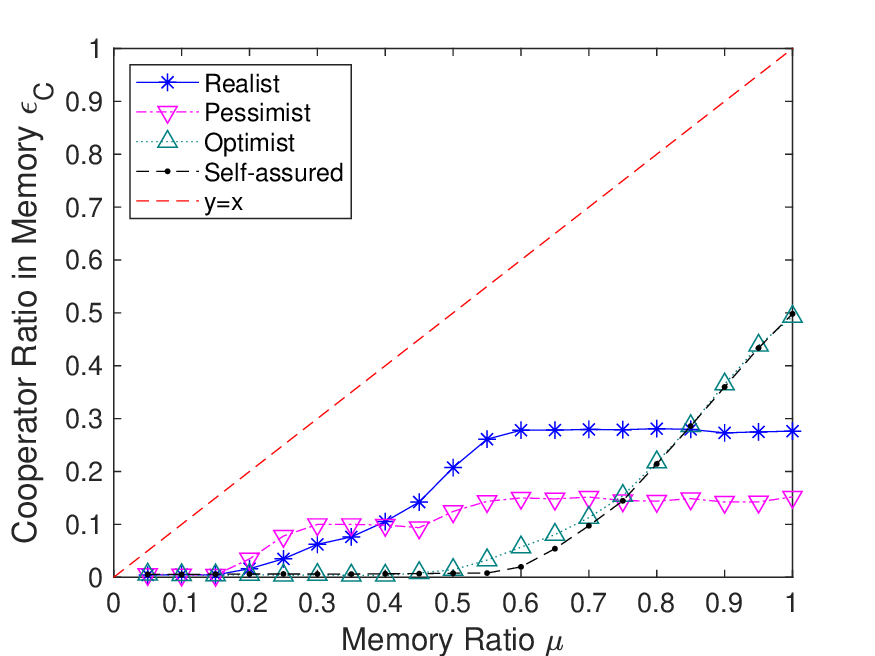}
		\label{fig:FC_memoryUsage_C}%
	}%
	\subfigure[Defectors in memory.]{%
		\includegraphics[width=0.45\columnwidth]%
			{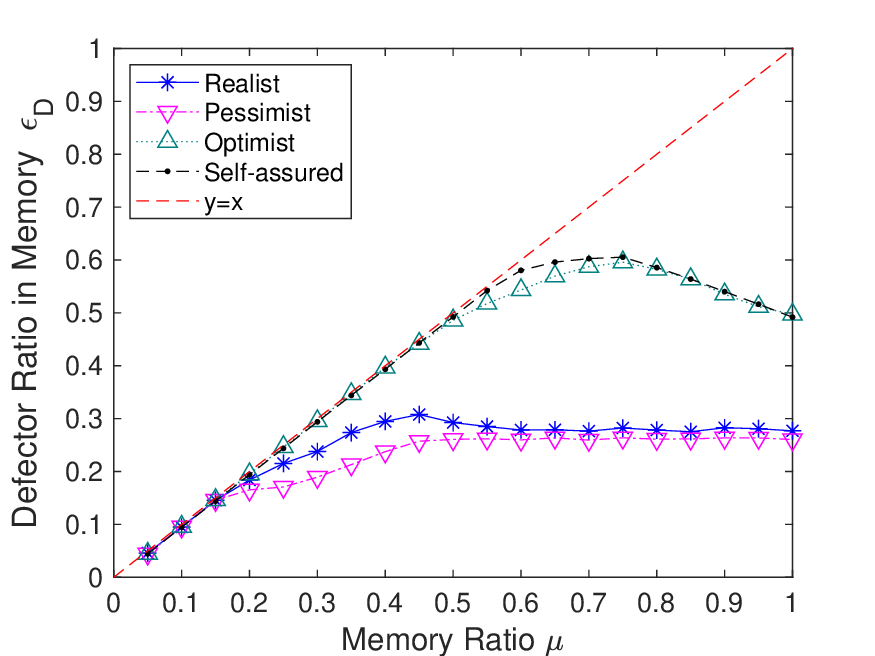}
		\label{fig:FC_memoryUsage_D}%
	}%
	\caption{
		Detailed analysis of FC.
		Line $x = y$ indicates the full memory usage in (d), (e) and (f).
	}%
	\label{fig:FC}%
\end{figure}

\subsubsection{Performance}

As expected, 
\reffig{fig:FC_PayoffRatio} shows that self-assured agents increase their performance as memory ratio increases.
Note that the self-assured curve in \reffig{fig:FC_PayoffRatio} and the FC curve in \reffig{fig:SA_All} are the same.

Interestingly, 
in \reffig{fig:FC_PayoffRatio},
we observe unexpected fluctuations in the performances of the optimists, pessimists and realists around $\mu = 0.2$.
Instead of increase, they decrease.
Note that, 
although agents receive recommendations,
their payoff ratio is worse than self-assured agents in this region.

For the values of $\mu > 0.2$,
performances start to increase again
but different dispositions have different trends.
Pessimists and realists are quick to recover and pass $\phi_{C} = 1$ threshold around $\mu = 0.25$.
They reach their peak values around $\mu = 0.45$ and stay there.
Optimists have a different path.
They cross $\phi_{C} = 1$ threshold around $\mu = 0.35$.
Their performance steadily increases and takes its peak value close to $\mu =1$.

\subsubsection{Number of evaluations}

In order to understand the strange behavior in performance,
we collect data on recommendations.

If an agent does not know the opponent,
it calls for recommendations.
If nobody in the 1-neigborhood knows the opponent,
then no response is obtained.
If the agent receives at least one response,
then it has to evaluate it.
\reffig{fig:FC_NORecommendations} plots
$\eta_{\times}$, 
the number of evaluations.
For all three dispositions, 
$\eta_{\times}$ increases as $\mu$ increases 
but it reaches to its maximum around $\mu = 0.2$.
Then different dispositions present different behaviors.
Pessimists keep the same level around $\eta_{\times} = 9.5 \times 10^{4}$ for $\mu > 0.3$.
For realists, the number of recommendation requests decreases till $\mu = 0.55$ and 
stays around $\eta_{\times} = 6.5 \times 10^{4}$.
For optimists, after its early peak value at $\mu = 0.15$,
the number of recommendation requests decreases almost to none.

\subsubsection{Evaluation accuracy}

In \reffig{fig:FC_AccuracyRatioEvaluation}, 
evaluation accuracy $\accuracyE$ values of cooperators are shown.
Since $50\%$ of the population is cooperators,
$\accuracyE = 0.5$ threshold is marked with a dashed line.
Note that the accuracies of all dispositions are above this line.

Initially, accuracy decreases as the memory of the agents' increases for all dispositions.
That is an unexpected result.
That is a result of cooperator-defector distribution in the memory when memory capacity is small.
There is a detailed explanation about this anomaly in the section~\ref{sec:anomaly0.2}.
Around $\mu =0.15$, dispositions start to deviate.
For realists, accuracy increases and reaches its maximum value of perfect accuracy, i.e. $\accuracyE = 1$, around $\mu = 0.4$.
The accuracy of pessimists increases slightly and 
stays just below $\accuracyE = 0.8$ for $\mu > 0.3$.

Behavior of optimists is the most difficult one to explain.
It keeps decreasing to just above $\accuracyE = 0.6$ as $\mu$ approaches to $0.4$,
then it has a sharp increase that reaches to above $\accuracyE = 0.9$ at $\mu = 0.55$,
and then has a smooth decrease back to $\accuracyE = 0.6$.

\section{Discussion}

Note that the way we set the experiment,
cooperators are of one type of disposition only.
We discuss each disposition separately.

\subsection{Optimists}

Once an agent plays with an agent,
it records this firsthand experience in its memory.
When the same agent match to play again,
it uses its memory
since it knows the agent.
Therefore,
it does not call for recommendations.

Optimists, by their nature, have an optimistic view of life.
One positive recommendation is good enough for them to play.
Since they play more,
they tend to know more people.
Since they know more people,
they ask less recommendation.
This explains the steady decrease of recommendation requests in \reffig{fig:FC_NORecommendations} for $0.2 < \mu < 1$.
For $0.2 < \mu < 0.4$ region in \reffig{fig:FC_AccuracyRatioEvaluation}, its accuracy keeps decreasing, too.
That is, it requests less recommendation and makes bad judgements.
Yet it knows enough defectors so that 
it can keep its payoff ratio increasing in this region.

\subsection{Pessimists}

Pessimists have the opposite strategy in playing. 
One single negative recommendation is enough for a pessimist not to play.
If it  does not play, 
then there will be no record of the opponent kept in its memory.
If the same opponent is matched to play again,
a pessimist has to ask recommendation once more.
This explains the high number of queries of recommendation even at $\mu = 1$, 
where there is enough memory to keep the entire population.

\subsection{Realists}

Realists are in the middle ground of optimists and pessimists.
They play more than pessimists but less than optimists.
In the region of $0.2 < \mu < 0.6$ of \reffig{fig:FC_NORecommendations},
similar to optimists,
they play with new agents and ``learn''.
From that point on,
their actions are similar to that of pessimists,
that is,
reject to play and keep asking the same agent over and over again.

\begin{figure}[htpb]
	\subfigure[False perception of defectors as cooperator.]{%
		\includegraphics[width=0.45\columnwidth]%
			{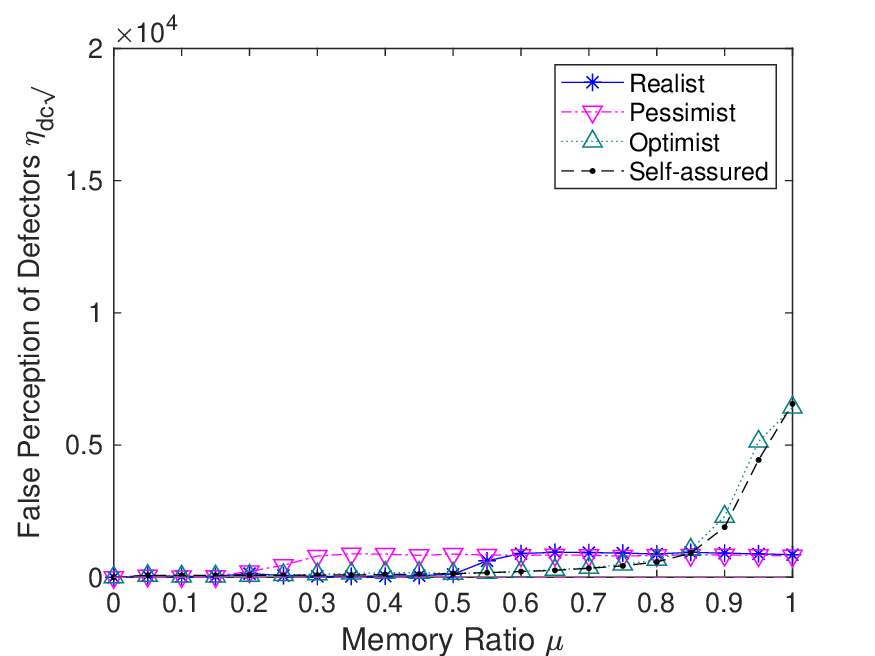}
		\label{fig:FC_percept_DasC}
	}%
	\quad
	\subfigure[False evaluation of defectors as cooperator.]{%
		\includegraphics[width=0.45\columnwidth]%
			{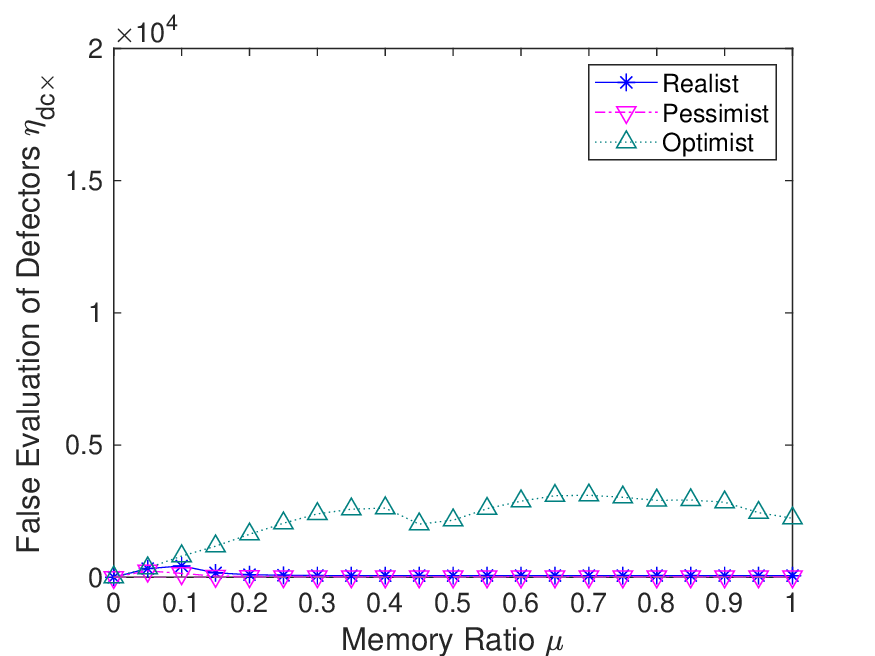}
		\label{fig:FC_eval_DasC}%
	}%
	\\ 
	\subfigure[False perception of cooperators as defector.]{%
		\includegraphics[width=0.45\columnwidth]%
			{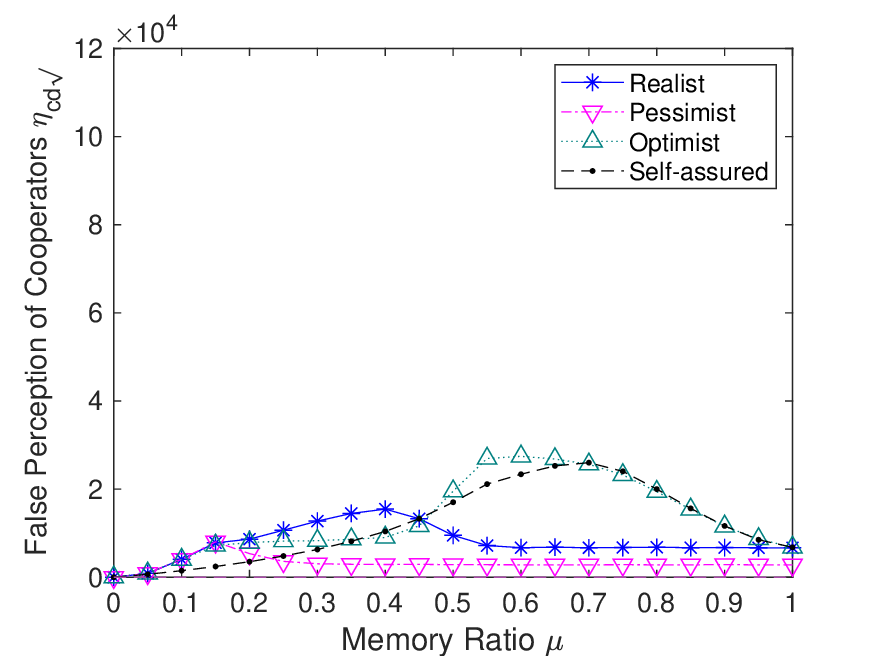}
		\label{fig:FC_percept_CasD}%
	}%
	\quad
	\subfigure[False evaluation of cooperators as defector.]{%
		\includegraphics[width=0.45\columnwidth]%
			{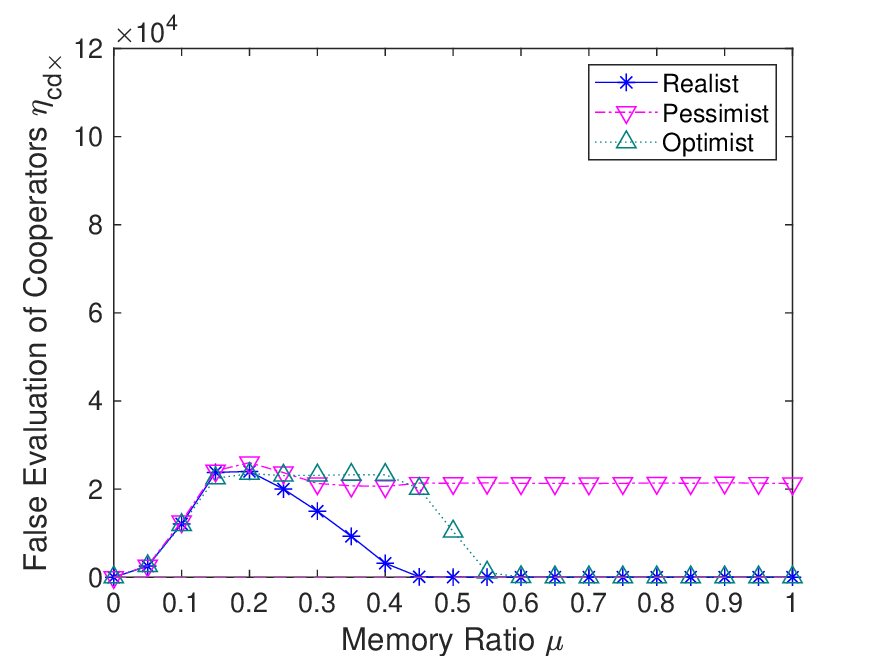}
		\label{fig:FC_eval_CasD}%
	}%
	\\ 
	\subfigure[Correct perception of cooperators.]{%
		\includegraphics[width=0.45\columnwidth]%
			{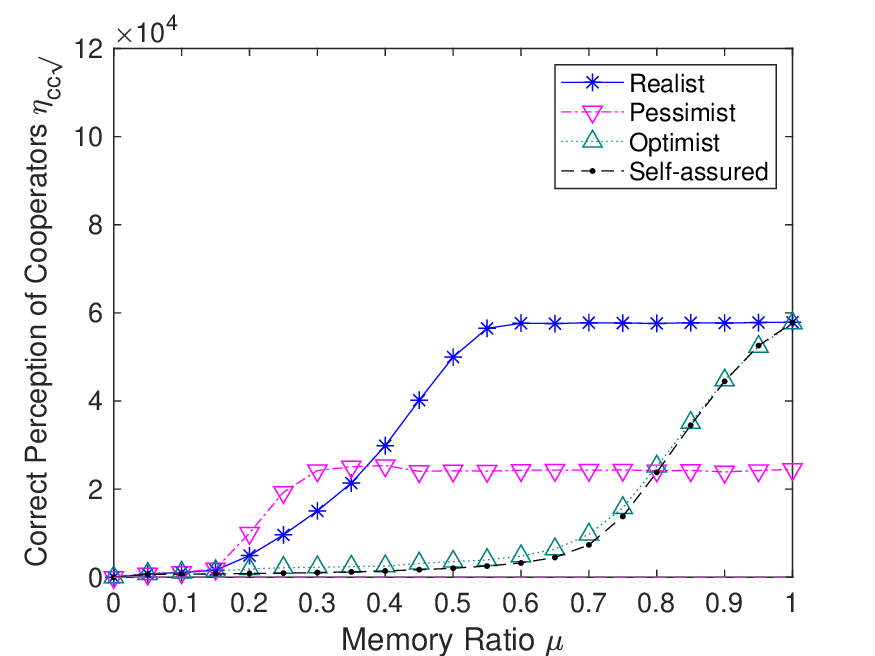}
		\label{fig:FC_percept_CasC}
	}%
	\quad
	\subfigure[Correct evaluation of cooperators.]{%
		\includegraphics[width=0.45\columnwidth]%
			{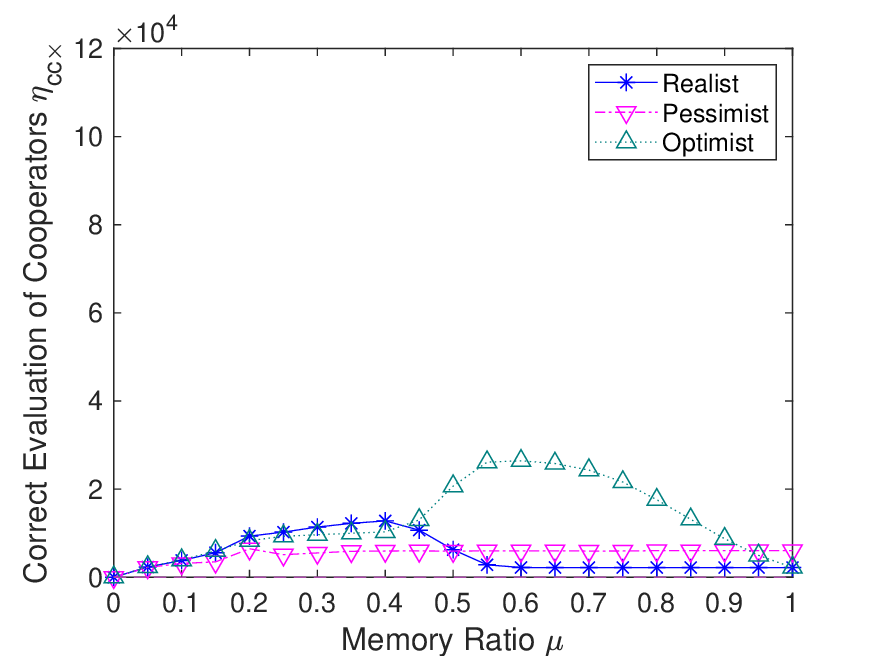}
		\label{fig:FC_eval_CasC}%
	}%
	\\ 
	\subfigure[Correct perception of defectors.]{%
		\includegraphics[width=0.45\columnwidth]%
			{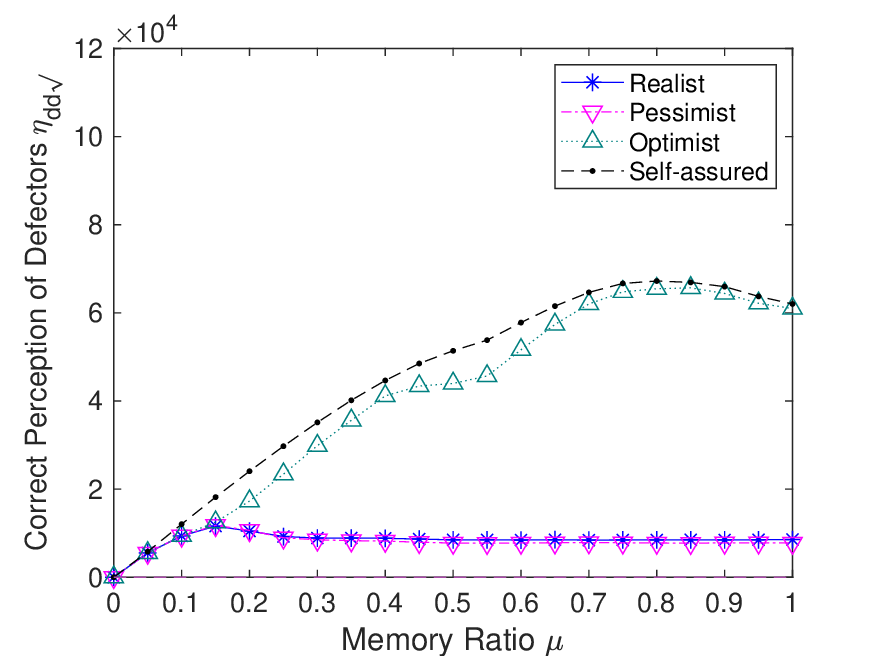}
		\label{fig:FC_percept_DasD}
	}%
	\quad
	\subfigure[Correct evaluation of defectors.]{%
		\includegraphics[width=0.45\columnwidth]%
			{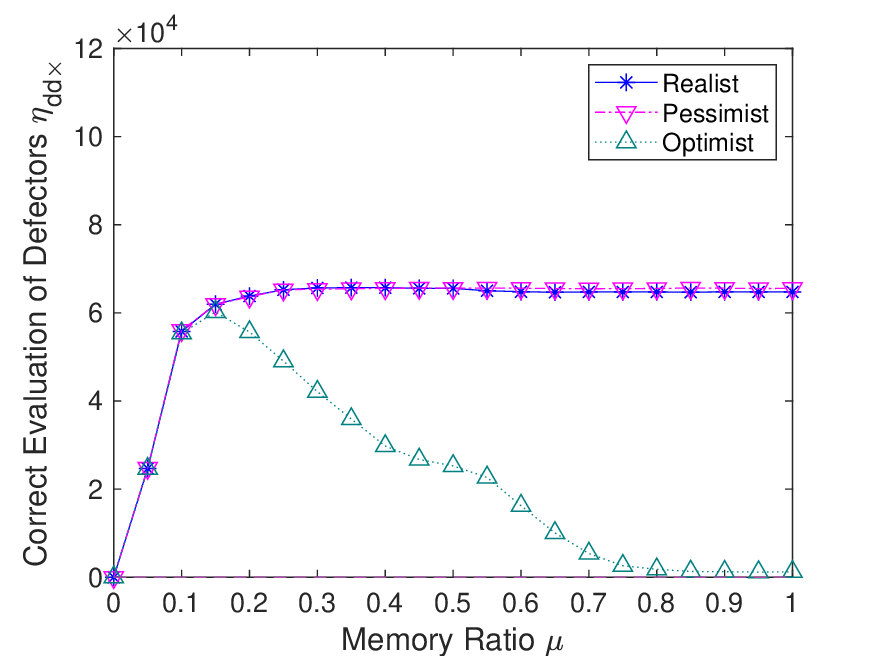}
		\label{fig:FC_eval_DasD}%
	}
	\caption{
		Breakdown of
		perceptions  and evaluations for FC.
		The first column is the perception of a known opponent.
		The second column is the evaluations of recommendations.
		Note that in (a) and (b), scale of $y$-axis is changed for better visibility.
	}%
	\label{fig:FC_eval_percept}%
\end{figure}

\subsection{Anomalies}

\subsubsection{Anomaly around $\mu = 0.2$}
\label{sec:anomaly0.2}

We observe a payoff drop in realists, optimists and pessimists 
around $\mu = 0.2$ in \reffig{fig:FC_PayoffRatio}.
In order to understand the surprising behavior in the region $0.15 < \mu < 0.25$,
we investigate how agents use their memories for small values of $\mu$.

First consider an ideal case of pure cooperators, 
i.e., $\rho = 1$, 
and defectors,
i.e., $\rho = 0$.
As usual any game played will be recorded in the memory.
Because of FC strategy, 
cooperative agents will be overwritten by the defectors.
Therefore, 
we expect that memory is full of defectors.
As $\mu$ gets closer to $50~\%$,
which is the percentage of defectors in the population,
we start to see cooperators in the memory.
For $\mu > 0.5$,
all the defectors are already in the memory.
Then the remaining memory is used for cooperators.

In our mostly cooperative,
i.e., $\rho = 0.9$, 
and mostly defector,
i.e., $\rho = 0.1$,
case,
agents act slightly different.
They store cooperators in their memories at early stages of $\mu < 0.5$.
In order to further investigate,
we need a few definitions.

Let $I_{i}^{c}$ and $I_{i}^{d}$ be the numbers of cooperators and defectors in the memory of agent $i$ at the end of simulations, 
respectively.
Let $I_{i}$ be the total number agents kept in the memory of $i$,
i.e.,
$I_{i} = I_{i}^{c} + I_{i}^{d}$.
As we will see later, 
agents may not use their memory fully,
i.e.,
$I_{i} \le M = \mu N$.
Then 
average memory usage, 
and
average ratio of perceived cooperators and
defectors in memory 
are defined, respectively, as
\begin{equation}
	\epsilon
	= 
	\frac{1}{|\mathcal{C}|}
		\sum_{i \in \mathcal{C}} \frac {I_{i}}{N},
	\quad \text{ and } \quad 
	\epsilon_{C}
	= 
	\frac{1}{|\mathcal{C}|}
		\sum_{i \in \mathcal{C}} \frac {I_{i}^{c}}{N},
	\quad 
	\epsilon_{D}
	= 
	\frac{1}{|\mathcal{C}|}
		\sum_{i \in \mathcal{C}} \frac {I_{i}^{d}}{N},
\end{equation}
where 
$\mathcal{C}$ is the set of cooperators.

\textbf{Payoff ratio.}
All dispositions use their full memory capacity for small values of $\mu <0.15$ in \reffig{fig:FC_memoryUsage_Total}.
Pessimists are the first ones to deviate using full memory around $\mu = 0.3$.
They are followed by realist around $\mu =0.55$.
Optimist and self-assured agents keep using their memory in full capacity in the entire range of $\mu$.

As expected, 
for small values of $\mu$,
memory is used for tracking defectors only in \reffig{fig:FC_memoryUsage_D}.
Since memory consists of defectors only,
there are no cooperators to recommend.
Therefore, 
any recommendation one gets would be a negative recommendation for defector as seen in \reffig{fig:FC_eval_DasD}.
Thus, 
even optimists reject to play.
That means cooperator agents do not get any points
while self-assured defectors do play and increase their points.
The result is the drop of payoff ratio for all dispositions around $\mu \sim 0.2$.
Note that the payoff ratio of self-assured cooperators continues to increase 
without affected by this
since they do not use recommendation.

As \reffig{fig:FC_memoryUsage_C} reveals,
both pessimist and realist stop using memory for defectors only starting around $\mu = 0.25$.
Since there are some cooperators in memory,
two things happen.
(i)~Realist and pessimist agents use their memory to correctly play with cooperators as seen in \reffig{fig:FC_percept_CasC}.
(ii)~Agents, that have cooperators in their memory, 
give positive recommendations for cooperators.
Some of them evaluated correctly \reffig{fig:FC_eval_CasC},
which increases the number of cooperators in the memory.
This leads to recovery in payoff ratio for pessimists and realists.

\textbf{Accuracy.}
Let's focus on the drop in the evaluation accuracy $\accuracyE$ 
observed around $\mu \sim 0.15$ in \reffig{fig:FC_AccuracyRatioEvaluation}.
Accuracy drops if an agent fails to correctly evaluate the recommendation.
Therefore, 
the drop should be the result of either failing to evaluate
cooperators as defector or defectors as cooperator
or both.
As seen in \reffig{fig:FC_eval_DasC},
the number $\nn{dc}$ of recommendations misjudged defectors as a cooperator is 
so small that a scale change in $y$-axes is required.
Therefore, 
accuracy drop should come from misjudgments of cooperators as a defector.
Increase in the number $\nn{cd}$ of misjudgments of cooperators as a defector in \reffig{fig:FC_eval_CasD} 
in the range of $0.05 < \mu <0.45$ causes the drop in the accuracy.
That can be clearly seen in $\accuracyE$ of realists in \reffig{fig:FC_AccuracyRatioEvaluation}.
Pessimist keeps making the same level of error staring from $\mu =0.2$ and its corresponding accuracy level stays at $\accuracyE = 0.8$.

\subsubsection{Anomaly around $0.55 < \mu < 0.7$}

Although one expects them to increase in \reffig{fig:FC_PayoffRatio},
payoff ratios of the optimist and self-assured agents remain constant in the range of $0.55 < \mu < 0.7$,
which calls for explanation.

In \reffig{fig:FC_eval_CasD},
recovery of optimist from false evaluation of cooperators as defector starts at $\mu = 0.4$.
It is completed at $\mu =0.6$.
But then another dynamic, 
i.e., mistakes in perception, 
takes place for both optimist and self-assured agents.
False perception of cooperators as defector in their memory reaches 
a peak around $\mu =0.6$ in \reffig{fig:FC_percept_CasD}.
This is observed as a flat segment of payoff in $0.5 < \mu < 0.7$ in \reffig{fig:FC_PayoffRatio}.

As $\mu$ approaches 1,
the behavior of optimists deserves special attention.
Since they are much more open to play
and use their entire memory,
they ``know'' more agents as $\mu$ increases.
Because of that,
they rely on their memory more.
\reffig{fig:FC_NORecommendations} confirms that 
they ask less recommendations as $\mu$ increases.
Although very small,
there is nonzero $\nn{dc}$ in \reffig{fig:FC_eval_DasC}.
For $\mu$ values close to 1,
the number $\numE$ of recommendations becomes so small that 
their ratio $\nn{dc} / \numE$ becomes visible as a slow decline in the accuracy in \reffig{fig:FC_AccuracyRatioEvaluation}.

\subsection{Possible extensions}

This work can be expanded in various ways.
%
Let's start with the parameters.
The model has several parameters
that one can change and observe the effects.
A set of important parameters are related to the population.
We considered $N = 100$ with $50\%$ cooperators.
One wants to try larger $N$ values as well as different percentages.
Among all the parameters of the model,
the payoff matrix is the most important one.
We used classical PD payoff matrix~\cite{%
	axelrod1981evolution}.
The model using different payoff matrices possibly produce different results 
but its investigation is left as future work.
For example, payoff matrices with negative entries,
as in the case of \cite{%
	Cetin2016},
would be an interesting possibility to investigate.
Another parameter to investigate is $\tau$,
which is the number of times two agents interact.
On the one hand,
one should not consider very small values.
For example,
for the extreme case of $\tau = 1$,
the model becomes trivial
since agents judge the opponents with a single interaction.
On the other hand,
very large values are not realistic in real life.
Except for our closed neighborhood,
the number of interactions that we made with an individual is not large.
In addition to $\tau = 30$ reported here,
we run tests for $\tau$ values of $10, 100, 200$ and
observed the ``anomaly around $\mu = 0.2$''.
See appendix~\refsec{sec:effectOfTau}.
Second, 
in our current model, 
the recommender agent gives its sincere perception as a recommendation.
That is, even a defector agent provides its genuine opinion.
This may not be the case in real life.
Third,
we considered homogenous agents.
One type of cooperators plays against one type of defectors.
In real life, there are always mixtures of all kinds.
In such heterogeneous environments are difficult to investigate but definitely much more realistic.
The fourth possible extension is related to the use of recommendation.
The original model of \cite{Cetin2014,Cetin2016}
has no recommendation.
In this work, we study the case where cooperators use recommendation and defectors do not.
There are two more cases to investigate:
cooperators do not but defectors do use recommendation and
both cooperators and defectors use it.
Fifth, 
our defectors refuse to play with perceived defectors.
Given a payoff matrix of $(S, P, R, T) = (0, 1, 3, 5)$,
a realist defector would always choose to play
since it has nothing to lose.
Finally,
in this work, we investigate the system at a certain time.
Its evolution in time would be interesting to investigate.

\section{Conclusion}

We investigated the Iterated Prisoner's Dilemma game where agents get recommendations if they do not know the opponent.
Having recommendation enables 
cooperators to beat the defectors
when they have  enough memory.

Although we expect better performance as memory capacity increases,
the performances of all dispositions drop around $\mu = 0.2$ region,
in which agents that do not get any recommendations perform better.
After that region performances recover.
Realists have the best performance while the optimists have the worst.

In this work, 
we report, in detail, strong cooperators and strong defectors with $\rho = 0.9$ and $\rho = 0.1$, respectively.
Although not reported here,
we also investigate mild cooperators and defectors
such as $\rho = 0.75$ and $\rho = 0.25$, and
we obtained similar results.
The source code is available at \url{https://github.com/zeynepcinar/IPDWithRec}.

\section*{Acknowledgements}
We would like to thank 
Uzay Cetin, 
Mursel Tasgin, 
and
Emre Aladag for constructive comments.
This work is partially supported by 
the Turkish Directorate of Strategy and Budget
under the TAM Project number 2007K12-873.

\appendix

\section{Effect of $\tau$}
\label{sec:effectOfTau}

\begin{figure}%
	\centering
	\subfigure[Optimist.]{%
		\includegraphics[width=0.5\columnwidth]%
			{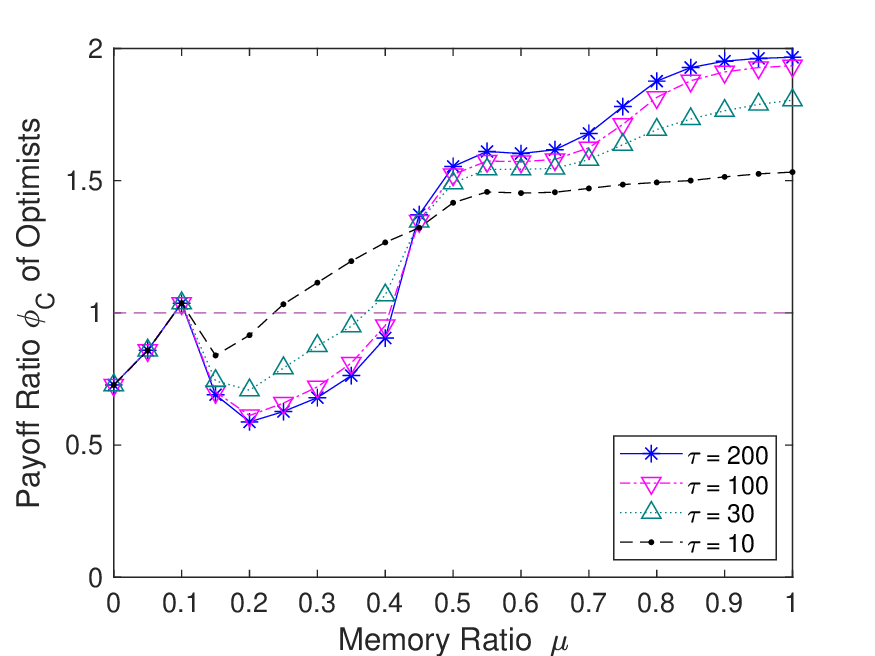}
		\label{fig:tau-Optimist}%
	}
	\subfigure[Realist.]{%
		\includegraphics[width=0.5\columnwidth]%
			{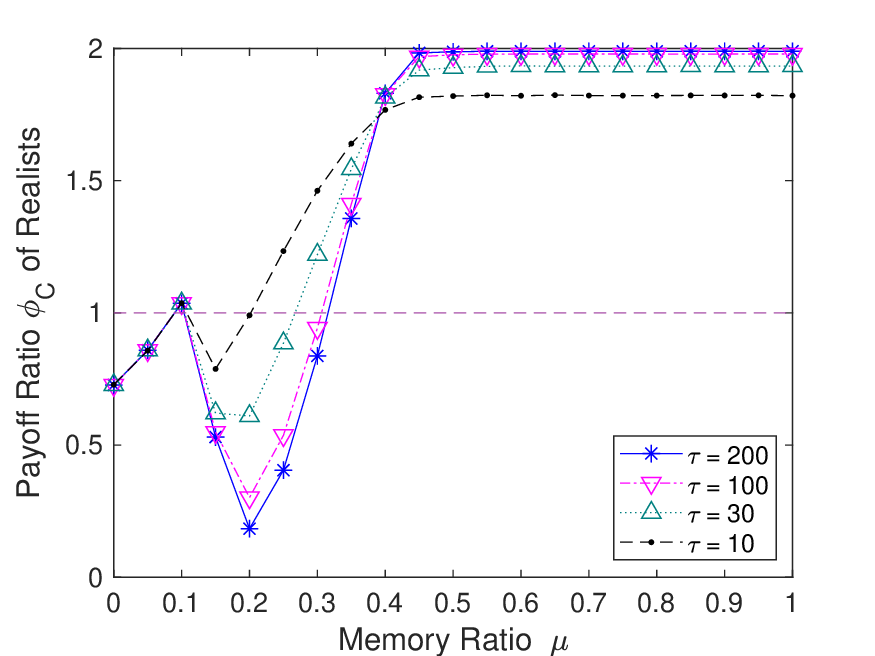}
		\label{fig:tau-Realist}%
	}%
	\\
	\subfigure[Pessimist.]{%
		\includegraphics[width=0.5\columnwidth]%
			{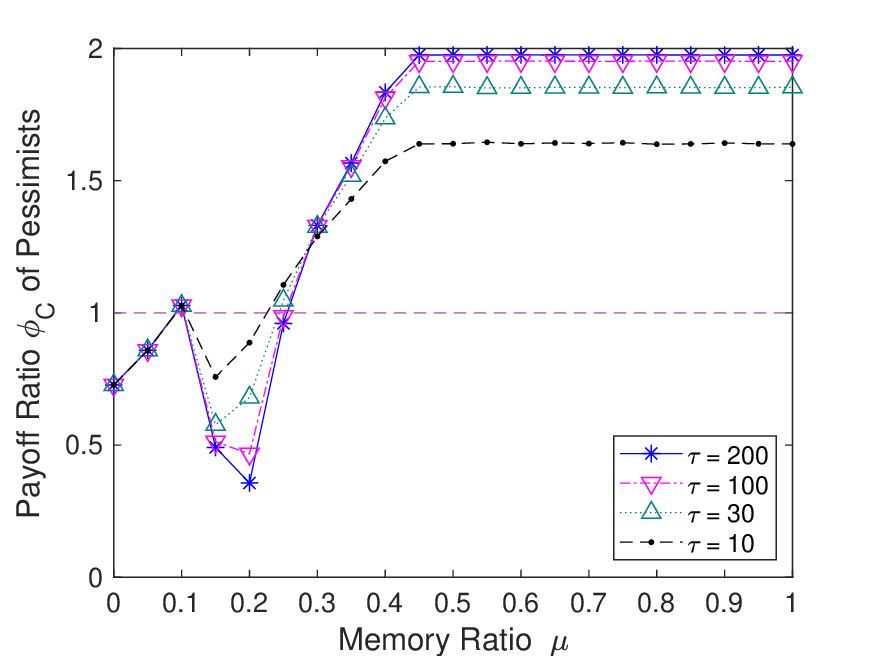}
		\label{fig:tau-Pessimist}%
	}%
	\subfigure[Self-Assured.]{%
		\includegraphics[width=0.5\columnwidth]%
			{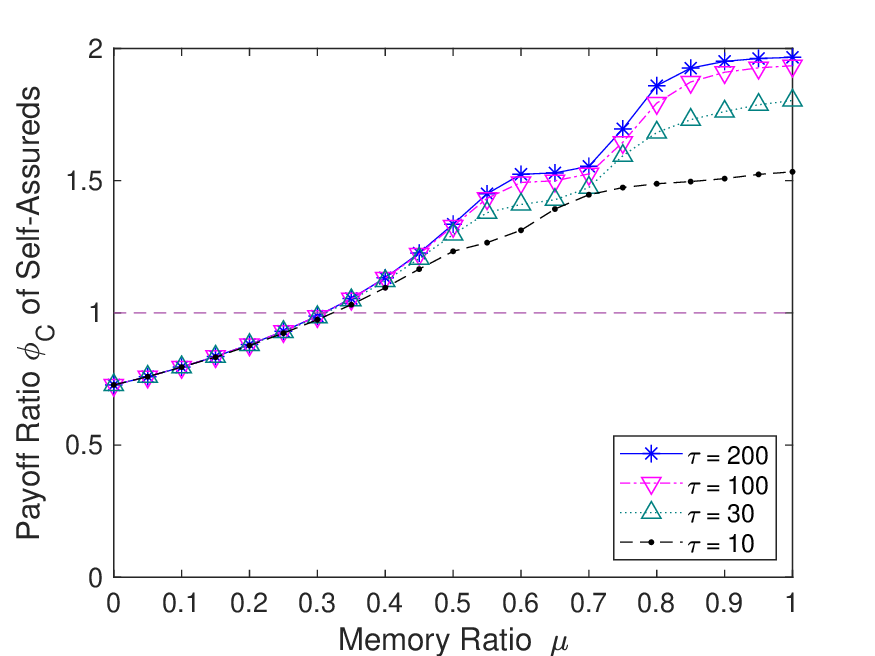}
		\label{fig:tau-SelfAssured}%
	}%
	\caption{Effect of $\tau$.}
	\label{fig:tau}%
\end{figure}

Note that $\tau$ is an appropriate time unit in this study.
One may investigate the behavior of the system for different values of $\tau$.
In addition to $\tau = 30$ reported in the manuscript,
we run tests for $\tau$ values of $10, 30, 100, 200$.
See \reffig{fig:tau}.
The ``anomaly around $\mu = 0.2$'' is observed in the plots of pessimist, optimist and realist agents.
As expected, the self-assured agents do not exhibit that behavior.
Further investigation of the effects of $\tau$ is left as future work.

\vspace*{3pt}   

\bibliographystyle{ieeetran}
\bibliography{IPDwRec-arXiv-r3}

\begin{thebibliography}{10}
\providecommand{\url}[1]{#1}
\csname url@samestyle\endcsname
\providecommand{\newblock}{\relax}
\providecommand{\bibinfo}[2]{#2}
\providecommand{\BIBentrySTDinterwordspacing}{\spaceskip=0pt\relax}
\providecommand{\BIBentryALTinterwordstretchfactor}{4}
\providecommand{\BIBentryALTinterwordspacing}{\spaceskip=\fontdimen2\font plus
\BIBentryALTinterwordstretchfactor\fontdimen3\font minus
  \fontdimen4\font\relax}
\providecommand{\BIBforeignlanguage}[2]{{%
\expandafter\ifx\csname l@#1\endcsname\relax
\typeout{** WARNING: IEEEtran.bst: No hyphenation pattern has been}%
\typeout{** loaded for the language `#1'. Using the pattern for}%
\typeout{** the default language instead.}%
\else
\language=\csname l@#1\endcsname
\fi
#2}}
\providecommand{\BIBdecl}{\relax}
\BIBdecl

\bibitem{bingol2008fame}
H.~Bingol, ``Fame emerges as a result of small memory,'' \emph{Physical Review
  E}, vol.~77, no.~3, p. 036118, 2008.

\bibitem{susan1991dynamics}
D.~Susan and J.~G. Holmes, ``The dynamics of interpersonal trust: Resolving
  uncertainty in the face of risk,'' \emph{Cooperation and Prosocial Behavior;
  Cambridge University Press: New York, NY, USA}, p. 190, 1991.

\bibitem{sherchan2013survey}
W.~Sherchan, S.~Nepal, and C.~Paris, ``A survey of trust in social networks,''
  \emph{ACM Computing Surveys (CSUR)}, vol.~45, no.~4, p.~47, 2013.

\bibitem{marsh1994optimism}
S.~Marsh, ``Optimism and pessimism in trust,'' in \emph{Proceedings of the
  Ibero-American Conference on Artificial Intelligence (IBERAMIA'94)}, 1994.

\bibitem{axelrod1981evolution}
R.~Axelrod and W.~D. Hamilton, ``The evolution of cooperation,''
  \emph{Science}, vol. 211, no. 4489, pp. 1390--1396, 1981.

\bibitem{Leas2016prisoners}
M.~Leas, E.~Dolson, R.~Annis, J.~Nahum, L.~Grabowski, and C.~Ofria, ``The
  prisoners dilemma, memory, and the early evolution of intelligence,'' in
  \emph{Proceedings of the Artificial Life Conference 2016 13}.\hskip 1em plus
  0.5em minus 0.4em\relax MIT Press, 2016, pp. 408--415.

\bibitem{nowak2006five}
M.~A. Nowak, ``Five rules for the evolution of cooperation,'' \emph{science},
  vol. 314, no. 5805, pp. 1560--1563, 2006.

\bibitem{chu2017win}
C.~Chu, J.~Liu, C.~Shen, J.~Jin, and L.~Shi, ``Win-stay-lose-learn promotes
  cooperation in the prisoner's dilemma game with voluntary participation,''
  \emph{PloS one}, vol.~12, no.~2, p. e0171680, 2017.

\bibitem{yamamoto2019effect}
H.~Yamamoto, I.~Okada, T.~Taguchi, and M.~Muto, ``Effect of voluntary
  participation on an alternating and a simultaneous prisoner's dilemma,''
  \emph{Physical Review E}, vol. 100, no.~3, p. 032304, 2019.

\bibitem{nowak1998dynamics}
M.~A. Nowak and K.~Sigmund, ``The dynamics of indirect reciprocity,''
  \emph{Journal of theoretical Biology}, vol. 194, no.~4, pp. 561--574, 1998.

\bibitem{nowak1998evolution}
------, ``Evolution of indirect reciprocity by image scoring,'' \emph{Nature},
  vol. 393, no. 6685, p. 573, 1998.

\bibitem{Cetin2014}
U.~Cetin and H.~O. Bingol, ``Iterated prisoners dilemma with limited
  attention,'' \emph{Condensed Matter Physics}, vol.~17, no.~3, p. 33001, 2014.

\bibitem{Cetin2016}
------, ``The dose of the threat makes the resistance for cooperation,''
  \emph{Advances in Complex Systems}, vol.~19, no.~08, p. 1650015, 2016.

\bibitem{stanley1993iterated}
E.~A. Stanley, D.~Ashlock, and M.~D. Smucker, ``Iterated prisoner's dilemma
  with choice and refusal of partners: Evolutionary results,'' in
  \emph{Proceedings of the Third European Conference on Advances in Artificial
  Life}.\hskip 1em plus 0.5em minus 0.4em\relax London, UK, UK:
  Springer-Verlag, 1995, pp. 490--502.

\bibitem{Marsh1994formalisingtrust}
S.~P. Marsh, ``Formalising trust as a computational concept,'' Department of
  Computing Science and Mathematics University of Stirling, Tech. Rep., 1994.

\bibitem{kiefhaber2013ranking}
R.~Kiefhaber, R.~Jahr, N.~Msadek, and T.~Ungerer, ``Ranking of direct trust,
  confidence, and reputation in an abstract system with unreliable
  components,'' in \emph{2013 IEEE 10th International Conference on Ubiquitous
  Intelligence and Computing and 2013 IEEE 10th International Conference on
  Autonomic and Trusted Computing}.\hskip 1em plus 0.5em minus 0.4em\relax
  IEEE, 2013, pp. 388--395.

\bibitem{jaynes2003probability}
E.~T. Jaynes, \emph{Probability theory: The logic of science}.\hskip 1em plus
  0.5em minus 0.4em\relax Cambridge University Press, 2003.

\bibitem{zabell1989rule}
S.~L. Zabell, ``The rule of succession,'' \emph{Erkenntnis}, vol.~31, no. 2-3,
  pp. 283--321, 1989.

\end{thebibliography}

\end{document}